# The Co-Evolution of Galaxies and Black Holes: Current Status and Future Prospects

**Timothy M. Heckman**

Center for Astrophysical Sciences

Department of Physics & Astronomy

Johns Hopkins University

**Abstract**

I begin by summarizing the evidence that there is a close relationship between the evolution of galaxies and supermassive black holes. They evidently share a common fuel source, and feedback from the black hole may be needed to suppress over-cooling in massive galaxies. I then review what we know about the co-evolution of galaxies and black holes in the modern universe ($z < 1$). We now have a good documentation of which black holes are growing (the lower mass ones), where they are growing (in the less massive early-type galaxies), and how this growth is related in a statistical sense to star formation in the central region of the galaxy. The opportunity in the next decade will be to use the new observatories to undertake ambitious programs of 3-D imaging spectroscopy of the stars and gas in order to understand the actual astrophysical processes that produce the demographics we observe. At high redshift ($z > 2$) the most massive black holes and the progenitors of the most massive galaxies are forming. Here, we currently have a tantalizing but fragmented view of their co-evolution. In the next decade the huge increase in sensitivity and discovery power of our observatories will enable us to analyze the large, complete samples we need to achieve robust and clear results

## 1. Introduction

As we look forward to the powerful suite of space- and ground-based observatories that will dominate astronomy in the coming decade, it is worthwhile to reflect upon the extraordinary state of astronomy here and now.

Capped by the spectacular results from NASA's WMAP mission, we now have precision measurements of the geometry, age, composition, and density-fluctuation power spectrum of the universe (Spergel et al. 2007). The predictions of inflationary cosmology and the ΛCDM paradigm have been confirmed, so that we also have a robust understanding of the development and evolution of the structure of the dark-matter backbone of the universe (e.g. Tegmark et al. 2004). This might suggest that our work is largely done. Paradoxically however, it is that minority-component of the mass-energy content of the universe with which we have the most experience by far - ordinary baryonic matter - whose cosmic evolution has proven to be most difficult to understand. Ordinary matter can interact dissipatively, radiate, cool, respond to pressure gradients, generate and respond to magnetic fields, undergo nuclear reactions, etc. This richness of physical processes leads directly to a corresponding richness of astrophysical phenomena that have inspired and challenged astronomers for centuries.

So, with the basic cosmological framework firmly in place, the frontier in the study of the formation and evolution of galaxies is to understand the complex "gastro-physics" of the gas-star-black hole cosmic ecosystem. While the crucial role of star formation and resulting feedback in the evolution of galaxies has been understood for decades, it has only been in the last few years that the role of supermassive black holes has been appreciated. The tight correlation between the mass of the black hole and the velocity dispersion and mass of the galactic bulge within which it resides (Ferrarese & Merritt 2000; Gebhardt et al. 2000; Marconi & Hunt 2003; Haring & Rix 2004) is compelling evidence for a close connection between the formation of the black hole and that of its host galaxy (e.g. Kauffmann & Haehnelt 2000; Granato et al. 2001).

Feedback from supermassive black holes is believed to be an essential process in the formation and evolution of galaxies. Galaxy formation models, which attempt to understand how galaxies form through gas cooling and condensation within a merging hierarchy of dark matter halos predict that massive galaxies should in general be surrounded by a substantial reservoir of gas. This gas should be cooling and forming stars. Over a Hubble time, these processes should have produced far more present-day galaxies with large stellar masses than are actually observed. Simply put, the mass spectrum of dark matter halos is a power-law, while the galaxy stellar-mass spectrum has an exponential cut-off (the standard Schechter function). Similarly, the cooling rate of hot gas in the haloes of massive galaxies today should be high enough to produce an easily-observable population of young stars (e.g. Kauffmann, White & Guiderdoni 1993). A great deal of theoretical and observational effort is now being expended on understanding why this simple expectation does not appear to be borne out by observations. Matter accreting onto central supermassive black holes can, in principle, provide a vast source of energy, and jets or outflows can provide mechanisms for transporting the energy to large enough radii to regulate cooling and star formation in the massive galaxies where such black holes live (e.g. Churazov et al. 2001; Croton et al. 2006; Bower et al. 2006; Hopkins et al. 2007; Di Matteo et al. 2007).

Thus, while we do have a rough conceptual framework for the co-evolution of galaxies and supermassive black holes, there is much that we do not know. How does the gas get into the galaxy bulge in the first place (mergers, secular processes, cold and/or hot accretion)? Once in the galaxy, how is gas transported from the bulge (radii of a few kpc) to the black hole's accretion disk (radii of tens of astronomical units)? What astrophysics sets the mass ratio of the gas turned into stars compared to the gas accreted by the black hole at the observed value of ~$10^3$ (Marconi & Hunt 2003; Haring & Rix 2004)? What about the magnitude and specific form(s) of the feedback from the supermassive black hole? What is the sequencing? Does the black hole or galaxy form first? Is this a once-in-a-lifetime transformative event in the life of a galaxy, or a more gradual, intermittent process?

My plan for the rest of the paper is as follows. I will begin on the firmest ground by discussing the modern universe ($z < 1$). We know quite a lot already about the overall landscape, but not much about the actual astrophysical processes that connect the fueling of the black hole to star formation in the galaxy bulge. Then I will discuss the early universe ($z > 2$), the epoch when the most massive black holes and galaxies were forming. There have been remarkable developments over the past few years, but the picture is still fragmented. Throughout, I will try to highlight how the next generation of observatories can be used to answer the questions posed above.

## 2. The Modern Universe – What Do We Know Now?

### 2.1. Overview

In the present-day universe there are two nearly independent modes of black hole activity. The first is associated with Seyfert galaxies and QSO's. Here, the black hole is radiating strongly (typically at greater than a few percent of the Eddington limit). This mode is the one primarily responsible for the growth in mass of the population of black holes today. Only the population of lower mass black holes (< $10^8$ solar masses) are growing at a significant rate today, and these black holes live in the lower mass galaxies (< $10^{11}$ solar masses). For this AGN mode, there is a strong link (at least in a statistical sense) between the growth of the black hole and star formation. Since star formation requires cold gas, this suggests that the fuel source for the black hole is also cold gas. The associated star formation can provide a significant source of feedback through supernova-driven winds.

The second mode (radio galaxies) is dominated by the most massive black holes living in the most massive galaxies (giant ellipticals). Here, the accretion rates are low and the accretion process is radiatively inefficient. Instead, energy is being extracted from the black hole in the form of highly collimated relativistic jets that produce nonthermal radio emission. The strong dependence of the radio luminosity function on the mass of the galaxy and the enhancement in the luminosity function for galaxies living at or near the centers of clusters can both be understood if the black hole is being fueled by the accretion of slowly cooling hot gas. Little star formation accompanies this mode, and this may be in part because the radio jets are heating the surrounding hot gas, decreasing its cooling rate, and suppressing star formation.

## *2.2. Seyfert Galaxies: The Population of Rapidly Growing Black Holes*

The relatively rapid growth of black holes that takes place in Seyfert nuclei and QSOs is accompanied by strong emission from the accretion disk and its environment. In principle this emission can swamp that of the host galaxy, making it extraordinarily difficult to determine the properties of the host galaxy and relate these to the growth of the black hole. Fortunately, nature has provided us with its own "coronagraph". The central black hole and its accretion disk are encircled by a dusty torus that is opaque to the ultraviolet and optical emission from the accretion disk (Krolik & Begelman 1988). When viewed nearer the polar axis of the torus the accretion disk is seen directly and the object is called a Type 1 Seyfert galaxy or (if sufficiently luminous) a QSO. When viewed from nearer the equatorial plane of the torus, the accretion disk can not be seen directly and the object is called a Type 2 Seyfert galaxy (or Type 2 QSO). In the Type 2 Seyferts the observed ultraviolet through near-infrared continuum is dominated by the host galaxy (e.g. Kauffmann et al. 2003c). The presence of the obscured AGN can be determined by the strong mid-infrared emission from the dusty torus, by the high-ionization narrow UV, optical, and IR emission-lines produced as the ionizing radiation from the accretion disk escapes along the polar axis of the torus and photoionizes gas in the surrounding kpc-scale Narrow-Line Region (NLR), and by hard X-rays that have passed directly through the torus (if the torus is Compton-thin).

Our largest and most homogeneous sample of 'local' ($z \sim 0.1$) Type 2 AGN and their host galaxies has come from the Sloan Digital Sky Survey (SDSS). These AGN were recognized by the emission of the NLR in the SDSS spectra (Kauffmann et al. 2003c; Hao et al. 2005). The AGN luminosity can be estimated from the strong [OIII]$\lambda$5007 emission-line and an empirically-derived bolometric correction (Heckman et al. 2004), while the black hole mass can be determined from the bulge velocity dispersion ($\sigma$) and the $M_{BH}$ vs. $\sigma$ relation in Tremaine et al. (2002). The SDSS spectra plus multicolor optical images and GALEX near-and far-ultraviolet images can be used to determine the fundamental properties of the host galaxies (star formation rate, stellar mass, velocity dispersion, structure, and morphology) and of the local clustering environment (Kauffmann et al. 2003a, 2004,2007a; Brinchmann et al. 2004; Li et al 2007; Wild et al. 2007; Reichard et al 2008; Martin et al 2007).

The first question we can ask is: Which black holes are growing? The SDSS sample of Type 2 AGN shows that most present-day accretion occurs onto lower mass black holes ($<10^8$ solar masses). In fact, the volume-averaged accretion rates of low mass black holes imply that this population is currently growing (doubling its mass) on a timescale that is comparable to the age of the Universe (Heckman et al. 2004). In contrast, the mass-doubling timescale is more than two orders of magnitude longer for the population of the most massive black holes ($>10^9$ solar masses). These dormant black holes evidently formed at early times, and once blazed as powerful QSOs. Thus, the strong cosmological evolution of the AGN luminosity function documented in recent X-ray surveys (e.g. Barger et al. 2005; Hasinger et al. 2005) is primarily driven by a decrease in the characteristic mass scale of actively accreting black holes.

The next question we can ask is: In which galaxies are black holes currently growing? It is now well-known that the present-day galaxy population consists of two families (e.g. Kauffmann et al. 2003b; Baldry et al 2004; Brinchmann et al. 2004; Schiminovich et al 2007). One family is disk-dominated, rich in cold gas, and experiencing significant rates of star-formation (the "blue" galaxies). The other family is bulge-dominated, depleted in cold gas, and experiencing little if any star formation (the "red" galaxies). The transition between these two families is remarkably abrupt: "red" galaxies dominate at stellar masses above $\sim 10^{10.5}$ solar masses and at stellar surface mass densities above $\sim 10^{8.5}$ solar masses/kpc$^2$ while "blue" galaxies dominate below.

It is very intriguing that a majority of galaxies that lie in "no-man's land" between the red and blue galaxy population (in a region in parameter space sometimes called the "green valley") are AGN (Martin et al. 2007; Kauffmann et al. 2007a). There are also systematic trends in host properties with AGN luminosity (Cid Fernandes et al. 2001; Kauffmann et al. 2003c; Kauffmann et al 2007a; Wild et al 2007). The higher the luminosity of the AGN, the younger is the age of the stellar population in the central-most few kpc, the higher is the likelihood that the galaxy has experienced a burst of star formation within the last Giga-year, and the greater is the amount of dust extinction towards the bulge (a proxy for the presence of cold gas).

In a volume-averaged sense, most present-day accretion onto black holes is taking place in galaxies with relatively young stellar populations, intermediate stellar masses, and high surface mass densities (Heckman et al. 2004; Wild et al. 2007).

These are hybrid galaxies with the structures of the red galaxies, the young stellar populations of the blue galaxies, and masses near the boundary between the two families. These results at z~0.1 are at least qualitatively consistent with what is known about AGN host galaxies at z~1 (Nandra et al. 2007; Rovilos et al. 2007; Rovilos & Georgakakis 2007).

The assessment above can be done in a more quantitative way. If we integrate over the SDSS volume, we find that the average ratio of star formation to black hole accretion in bulge-dominated galaxies is ~$10^3$. This is in remarkable agreement with the observed ratio of stellar mass to black hole mass in nearby galaxy bulges. For the population of bulge-dominated galaxies, the volume-averaged mean growth time of the black hole (due to accretion) and the mean growth time of the surrounding bulge (due to star formation) are very similar. After volume averaging, the growth of black holes through accretion and the growth of bulges through star formation are thus related at the present time in the same way that they have been related, on average, throughout cosmic history.

Thus, it seems likely that the processes that established the tight correlation between bulge mass and black hole mass are still operating in low redshift AGN. These results also suggest a picture in which star formation and black hole growth have been moving steadily and in parallel to lower and lower mass scales since z~2 ("downsizing"). This is consistent with the cosmic evolution in the AGN luminosity function and inferred black hole mass function (Ueda et al. 2003; Barger et al. 2005; Hasinger et al. 2005; Fine et al. 2006; Vestergaard et al. 2008).

But by what process is the evolution of the black hole and galaxy linked? Some interesting clues emerged from Kauffmann et al. (2007a), who examined the relationship between the growth of the black hole and star-formation in the both the inner (bulge) and outer (disk) regions using SDSS and GALEX. They showed that galaxies with red (old) disks almost never have a young bulge or a strong AGN. Galaxies with blue (young) disks have bulges and black holes that span a wide range in age and accretion rate, while galaxies with young bulges and strongly accreting black holes almost always have blue outer disks. Their suggested scenario is one in which the source of gas that builds the bulge and black hole is a reservoir of cold gas in the disk. The presence of this gas is a necessary, but not sufficient condition for bulge and black hole growth. Some mechanism must transport this gas inwards in a time variable way.

What is this mechanism? In most of the popular models for the co-evolution of black holes and galaxies, the dominant phase of black hole growth occurs in the aftermath of a major merger between two gas-rich disk galaxies (e.g. Di Matteo et al 2007; Hopkins et al. 2007). This growth coincides with or closely follows a central burst of star-formation. However, at least in the modern universe (z < 1), a strong connection between major mergers and black hole growth has not been es-tablished. Simple by-eye examination of images of the host galaxies of the most rapidly-growing black holes in the SDSS sample of Type 2 AGN shows that most are normal or mildly perturbed systems (Kauffmann et al. 2003c). Similar results have been reported for Type 2 AGN at z~1 (Grogin et al. 2005; Pierce et al. 2007).

This result has been quantified by Reichard et al. (2008) who measured the "lopsidedness" of the stellar distribution in the host galaxies of low-redshift Type 2 AGN in the SDSS. They find that the more rapidly growing black holes live in galaxies that are more lopsided on-average. However, the typical lopsidedness of the hosts of even the most powerful AGN is mild (~10%). Moreover, Li et al. (2007), find no evidence for an excess of close companion galaxies near powerful Type 2 AGN in the SDSS. Li et al. (2007) and Reichard et al (2008) suggest that one way to reconcile their two results is by a two-stage process associated with a minor merger (e.g. Mihos & Hernquist 1994). During the first passage of the companion, an initial inflow of gas occurs that fuels star-formation. The growth of the black hole occurs only later in the end stages when the merger is complete (and during or just after a second and more intense episode of star formation).

This would be consistent with observed link between black hole growth and star formation described above, but there would be an offset in time. This scenario would also be qualitatively consistent with the results of Kauffmann et al. (2003c) and Wild et al. (2007) who found an unusually large fraction of powerful Type 2 AGN in "post-starburst" systems. However, these post-starbursts are still only a minority (~20%) of powerful Type 2 AGN. Wild et al. (2007) concluded that the merger-starburst-AGN scenario was a significant – but not the dominant – channel for the growth of black holes today.

As noted in the introduction, feedback from the AGN is believed to be an important (some would say essential) process in the evolution of massive galaxies. Some of the most successful models assume that a significant amount of the rest-mass energy of the matter accreted by the black hole is available to drive a powerful galactic wind (e.g. Hopkins et al. 2007, Di Matteo et al. 2007). To date, there is not clear evidence in the local universe that such processes typically operate in the generic radio-quiet AGN that are the majority population (I will discuss the radio-loud AGN below).

While outflows in radio-quiet AGN are commonly detected in the form of blueshifted absorption-lines (e.g. Weymann et al 1981), their energetics are very poorly constrained because we do not know their size-scale. In the one case where the dimensions of the outflow have been directly determined, the estimated outflow rate in kinetic energy is only of-order $10^{-6}$ $L_{Bol}$ (Krongold et al. 2007). On the other hand, Tremonti et al. (2007) have discovered outflows in extreme post-starburst galaxies at intermediate redshifts. The high outflow velocites (~$10^3$ km/sec) strongly suggest these flows were powered by a past AGN episode. If this AGN episode happened as long-ago as the starburst (a few hundred Myr), the implied size scale of the outflow is several hundred kpc, and the associated outflow rates are very high. Finally, small (kpc-scale) and relatively low-power radio jets are seen in many Seyfert galaxies, but they do not appear to be having a significant effect on the bulk of the host galaxy's ISM (Veilleux, Cecil, & Bland-Hawthorn 2005).

It is also important here to emphasize that since there is strong link between the growth of black holes and star formation, feedback in the form of the kinetic energy supplied by supernovae and stellar winds from massive stars will be available even if the black hole doesn't help at all. As an illustration, the formation of a $10^8$ solar mass black hole would be (eventually) accompanied by the formation of $\sim 10^{11}$ solar masses in stars, resulting in $\sim 10^9$ supernovae and $\sim 10^{60}$ ergs in kinetic energy. This is equivalent to $\sim 1\%$ of the rest-mass energy of the black hole, and is sufficient (in principle) to expel the entire ISM of such a galaxy.

## *2.3. Radio Galaxies: Homes of the Most Massive Black Holes*

Unlike the case of the Seyfert galaxies and QSOs, typical radio galaxies in the present-day universe appear to be accreting at a highly sub-Eddington rate, and in a radiatively inefficient mode (e.g Allen et al. 2006). Most of the energy extracted from the black hole appears in the form of highly collimated relativistic outflows of radio-emitting plasma.

For low-power radio galaxies, the large-scale radio emission takes the form of oppositely-directed twin jets whose surface-brightness steadily declines with increasing distance from the nucleus (Fanaroff & Riley 1974 – "FR"). These "FR I" sources totally dominate the local radio galaxy population by number. At very high radio power the morphology of radio sources is strikingly different: it is dominated by bright "hot spots" located on opposite sides of the galaxy at the two outer edges of the radio source. These "FR II" radio galaxies are so rare in the local universe that there are few of them in the SDSS main galaxy sample. However, their comoving density evolves strongly with redshift (e.g. Dunlop & Peacock 1990). Accordingly, I will discuss the FR I radio galaxies here, and their more powerful kin in the section below on the early universe.

A cross-match of the SDSS main galaxy sample and the FIRST and NVSS radio catalogs, yields a sample of several thousand local (z ~ 0.1) low-power radio galaxies (Best et al. 2005a). Best et al. (2005b) and Kauffmann et al. (2007b) showed that this class of AGN is almost totally disjoint from the Seyfert/QSO population described above. The properties of the radio galaxies differ in several strong and systematic ways from the Type 2 Seyfert galaxies described in the preceding section. The radio galaxies are selectively drawn from the population of the most massive galaxies (quantifying the well-known result that they are typically giant elliptical galaxies). More precisely, Best et al. found that the probability that a galaxy was a radio-loud AGN (at fixed radio power) increased as the total stellar mass of the galaxy to the ~2.5 power and as the galaxy velocity dispersion to the ~6.5 power. The radio galaxies also had the old stellar populations and highly concentrated structures of normal giant elliptical galaxies.

The strong mass-dependence of the radio luminosity function can be understood as reflecting the mass-dependence of the cooling rate of hot gas in elliptical galaxies (Best et a. 2005b). In fact, Chandra high- resolution X-ray imaging spectroscopy of the hot gas in the centers of the nearest such radio galaxies shows that the radio jet power scales directly with the estimated rate of Bondi accretion in these systems (Allen et al, 2006). Further support for this fueling scenario comes from the enhanced probability of radio emission from the brightest galaxies at the centers of groups and clusters of galaxies – just the locations where the cooling rates will be unusually high (Best et al. 2007).

While – as discussed above - the nature of feedback operating in the Seyfert/QSO mode is still unclear, there is now quite compelling evidence for feedback provided by radio sources. The most direct evidence comes from observations of the cavities evacuated in the hot gas in the cores of galaxy clusters by radio sources (e.g. Fabian et al. 2006; Birzan et al. 2004, McNamara & Nulsen. 2007). The sizes of the cavities and the measured gas pressure allows the amount of P$\Delta$V work done by the radio source to be calculated, while the sound crossing time of the cavity gives a characteristic time scale. This then allows a rough estimate to be made of the time-averaged rate of energy transported by the radio jets.

Best et al. (2006) used this approach to derive the scaling between the observed radio luminosity of the jet and its rate of energy transport. Combining this scaling relation with the mass-dependent radio luminosity function in Best et al. (2005b) they were able to show that the time-averaged mass-dependent heating rate due to radio jets roughly matches the measured average mass-dependent radiative cooling rates (X-ray luminosities) in elliptical galaxies.

While these low-power FR I sources do not evolve with look-back time as strongly as the FR II sources discussed below, work by Sadler et al. (2007) shows that their co-moving density roughly doubles between $z \sim 0$ and $z \sim 0.7$. The associated heating rate per co-moving volume would likewise have been larger then, while the total amount of stellar mass in elliptical galaxies was smaller (Bell et al. 2004; Faber et al. 2007). Thus, this form of feedback would have more important (more ergs per gram) than at present.

## 3. The Modern Universe: Prospects for the Next Decade

From the summary above, it appears that we have a pretty clear overall map of the basic landscape in the local universe: we know which black holes are growing and in what kinds of galaxies. Having said that, it is also clear that we have a very incomplete understanding of the actual astrophysical processes at work (we have good cartoons and slogans!). In what follows, I will highlight a few issues where I think the combination of major new facilities will allow us to make dramatic discoveries in the next decade.

### *3.1. The Universe at Redshift One*

At a redshift of one, the global rates of both star-formation and black hole growth were about an order-of-magnitude larger than today (e.g. Marconi et al. 2004). This is very intriguing because (from what we have seen so far) the universe at $z \sim 1$ does not look radically different from that at $z \sim 0$. The familiar galaxies that define the Hubble sequence are in place, and the scaling relations and building blocks that define the basic structures of galaxies have not evolved strongly (e.g. Barden et al. 2005; Jogee et al. 2004). As summarized above, the growth of black holes seems to occur in galaxies that are at least qualitatively similar to those in which black holes grow today. The relationship between black hole mass and galaxy velocity dispersion has apparently evolved only weakly if at all (Salviander et al. 2007). So the $z \sim 1$ universe looks sort of like the $z \sim 0$ universe on steroids. The most important difference presumably is the higher overall amount of cold gas in (and/or accretion rate onto) galaxies at $z \sim 1$.

With the advent of major new wide field multi-object spectrographs on 8 and 10 meter-class telescopes it would be feasible in the next decade to undertake the rough equivalent of the SDSS at $z \sim 0.5$ to 1. I'd like to highlight WFMOS - the Wide Field Multi-Object Spectrograph which is planned as a collaboration between the Gemini and Subaru Observatories. It would consist of 4500 fibers that could be deployed over a field of view of 1.5 degrees. These fibers would feed optical spectrographs with resolutions of $R \sim 4000$ ($\sigma \sim 30$ km/s).

### *3.2. The Fueling of Black Holes*

We now know that there is a strong link (at least in a statistical sense) between star formation in the innermost several kpc and the growth of black holes. Perhaps the most important unanswered question about AGN is how some small fraction of the cold gas supply being used to form stars is transported over many orders-of-magnitude in radius to the accretion disk around the black hole.

It is generally supposed that at least the early stages of this is driven by the angular momentum transport provided by a non-axisymmetric gravitational potential together with the loss of binding energy through dissipation and radiative cooling of the gas orbiting in such a system. Bars are the most well-known examples of such non-axisymmetric structures, but unfortunately, observational evidence linking the growth of black holes to the presence of either large-scale or nuclear bars has been inconclusive at best (e.g. Mulchaey & Regan 1997; Knapen, Shlosman, & Peletier 2000; Laine et al. 2002; Erwin & Sparke 2002). Oval distortions and even global spiral arms can also drive inflows (Kormendy & Kennicutt 2004 and references therein). Simoes Lopes et al. (2007) used HST imaging to show that all 65 AGN host galaxies in their sample have dust (and hence gas) in the central-most few hundred parsecs. Kinematic mapping of several such systems containing low-luminosity AGN shows the streaming of gas inward along the spiral arms seen in the dust (Fathi et al. 2006; Storchi-Bergmann et al. 2007; Riffel et al. 2008).

The phase of gas that will dominate by mass on such circum-nuclear scales will be the molecular gas traced at mm-wavelengths. Moreover, this dense and cool gas will be less subject to the sorts of AGN-driven flows often seen in the hotter and less dense gas (Veilleux, Cecil, & Bland-Hawthorn 2005). Exciting steps in this direction are being taken with current facilities (e.g. Lindt-Krieg et al. 2007). The capabilities of ALMA will be superbly matched to this problem for nearby AGN. The AGN lifetime may well be shorter than the characteristic flow times on these scales (e.g. Martini et al. 2003). If so, it will be essential to investigate the amount of potential fuel (molecular gas), its structure, and its dynamics for a large and complete sample of galaxies (e.g. without regard to the presence of an AGN). The huge increase in sensitivity provided by ALMA would make this feasible.

A related problem is that we do not know the astrophysics that sets the ratio of star-formation to black hole accretion to a time-averaged value of $\sim 10^3$. To tackle this problem we need to combine the information about the rate of inward gas flows as a function of radius (above) with a mapping of the radial distribution and time-dependence of star-formation. This can be provided by mapping of the stellar population and ionized gas in these regions (e.g. Gonzalez Delgado et al; 2004; Davies et al. 2007) using the new capabilities provided by JWST, GSMT, and ELT.

These capabilities will also be essential to probe what's going on at even smaller (few pc) scales. Here we enter a new regime where the gravitational potential of the black hole itself starts to dominates the dynamics. This transition radius is given by $R = GM_{BH}/\sigma^2$ where $\sigma$ is the galaxy velocity dispersion. Using the observed relation between $\sigma$ and $M_{BH}$ (Tremaine et al. 2002; Ferrerese & Merritt 2000), this can be written as R ~ 12 $(M_{BH}/10^8)^{1/2}$ pc. For the nearest AGN this has an angular scale of-order 0.1 arcsec. These scales are also where we approach the domain of the obscuring torus.

Of course, we have a lot of information about this region in the Galactic Center (e.g. Genzel & Karas 2007), but our own supermassive black hole there is currently quiescent. Observations of water masers using VLBI techniques provide our highest resolution maps of this region in nearby AGN (e.g. Moran et al. 2007), but only a minority of AGN have such masing regions (Braatz et al. 2004). With existing mid-IR interferometers and AO-fed ground-based near-IR spectrographs it is now just possible to resolve the hot dust emission from this region (Tristram et al. 2007) and map the kinematics of the hot molecular hydrogen (Davies et al. 2006; Hicks & Malkan 2008). In the next decade, JWST, ELT, and GSMT will make it possible to make detailed maps of the gas and dust on these small scales for complete samples of the nearest AGN.

### 3.3. AGN Feedback

The powerful capabilities for X-ray spectroscopy provided by the proposed International X-ray Observatory (IXO) will give us revolutionary insights into AGN-driven feedback. With an imaging X-ray calorimeter it will be possible for the first time to actually make spatially-resolved maps of the detailed kinematics of the hot gas that is being accelerated and heated by radio sources in clusters and giant elliptical galaxies. The combination of high spectral resolution and sensitivity will also make it possible to conduct a major campaign of time-domain studies of the AGN-driven outflows traced through their blue-shifted X-ray absorption-lines. Such studies are the best way to determine the size-scales of these outflows, and hence the outflow rates of mass and energy carried by them.

## 4. The Early Universe

### 4.1. Overview

In summarizing above what we know about the modern universe, I emphasized results from large, homogeneously selected, and complete samples. I also empha-sized the importance of spectroscopy. Such an approach has allowed us to decisively resolve some long-standing controversies, quantify long-known or long-suspected qualitative results, and discover the unexpected. Unfortunately, at high-redshift this approach is essentially impossible with our current capabilities.

It's obvious that at high redshift the host galaxies of AGN are faint, and so it is not currently possible to observe very large samples. A less obvious but equally important problem is that it is very difficult with the present data to robustly characterize the basic properties of both the AGN (e.g. bolometric luminosity, black hole mass, Eddington ratio) and those of its host galaxy (e.g. stellar mass, star formation rate, velocity dispersion, structure/morphology). We have a reasonable handle on the AGN properties in high-z QSOs, but then have very limited information about their host galaxies. As at low redshift, it is easier to study the properties of the host galaxies in Type 2 AGN. The main techniques for finding Type 2 AGN at high-redshift are through observations in the rest-frame mid-IR or in the hard X-ray band, supplemented by radio continuum observations. Unfortunately, with only this limited coverage of the full spectral energy distribution, it is often difficult to cleanly separate out AGN from powerful starbursts.

Here's one way of thinking about the problem. We know that star formation and black hole growth are coupled, and that in a time-averaged sense the ratio of these two rates is of-order $10^3$ (e.g. Marconi et al. 2004). Models suggest that this ratio is more like $10^2$ during the phase when the black hole growth rate is maximized (e.g. Hopkins et al. 2007). The ratio of bolometric luminosity (stars/black hole) corresponding to the two ratios above are ~10 and ~1 respectively (assuming a radiative efficiency of 10% for black hole accretion and assuming a Kroupa IMF for the star formation). Thus, we expect both phenomena to be energetically significant in high-redshift AGN. Sorting out the luminosity attributable to the AGN (measuring the black hole accretion rate) and to the young stars (determining the star-formation rate) is not straightforward with the limited available data. Even assuming we can determine the AGN bolometric luminosity, we usually have only a lower limit to the black hole mass in Type 2 AGN (by assuming that the Eddington limit is obeyed).

### *4.2. Which Came First?*

The tight relationship at z ~ 0 between the mass of the black hole and the mass and velocity dispersion of the galaxy bulge within which it lives is the result of a time integral over the history of the universe. It does not however tell us that the two formation processes must occur simultaneously in any given galaxy. If not, which comes first: the black hole or the galaxy?

There have been a number of attempts to determine the relationship between the galaxy mass (or velocity dispersion) and black hole mass at high-redshift. Peng et al. (2006) used HST images of eleven QSO host galaxies at z ~ 2 to estimate luminosities of the bulges. They estimated the black hole masses using the "photoionization method". This is based on the scaling relation seen in local Seyfert nuclei between black hole mass (determined via reverberation mapping), AGN ionizing luminosity, and the width of the emission-lines in the Broad Line Region (Kaspi et al. 2005). Peng et al find that the ratio of black hole to bulge mass is about four times larger at z ~ 2 than at present (e.g. the black holes form first). McLure et al. (2006) find a similar result using a matched sample of radio galaxies and radio-loud QSOs.

In contrast, Shields et al. (2003) examined a sample of QSOs at redshifts up to ~3. They used the velocity dispersion of the emission-lines in the NLR as a proxy for the stellar velocity dispersion (Nelson & Whittle 1996; Greene & Ho 2005). The black hole masses were estimated using the above photoionization method. They found no evidence that the relation between black hole mass and galaxy velocity dispersion had evolved strongly between z ~ 0 to 3. Borys et al. (2005) point out that if the black holes detected in sub-mm galaxies via their hard X-ray emission are accreting at the Eddington rate, their luminosites would imply that they have masses more than an order-of-magnitude below that of black holes today in galaxies with similar mass (e.g. the galaxy would come before the black hole).

Overall, the situation regarding the phasing of black hole formation at high-z is thus unclear. Hopkins et al (2006) use integral constraints on the co-moving mass density of black holes to argue that there can not be a strong evolution in the ratio of stellar and black hole mass in galaxies. Small changes will be difficult to detect, especially when the ferocious systematic effects discussed by Lauer et al. (2007) are considered. Lauer et al. argue that a robust attack on this problem requires both an accurate measurement of the "scatter function" in the $M_{BH}$ vs. $M_{gal}$ or vs. $\sigma$ relations, and samples at high and low redshift that have been selected (and investigated) in exactly the same way using precisely defined and objective criteria. This task will be far easier using the capabilities of JWST, GSMT, and ELT to characterize the salient properties of large samples of galaxies at high-redshift.

## *4.3. Where are Black Holes and Galaxies Growing?*

I'd like now to briefly discuss observations of the connections between black hole growth and star formation at high redshift by looking for signs of both in a given population of galaxies. Because of the multitude of ways in which both high-z AGN and high-z galaxies are selected and investigated, I am strongly re-minded of the fable of the blind men and the elephant (whom the blind men variously describe as a tree, a large leaf, a rope, and a wall depending upon their particular set of "observations").

The largest sample of high-z AGN by far are QSOs. Observations of small samples of high-z QSOs that were selected based on their relative brightness in the rest-frame far-IR (indicating strong dust emission) show that these have very large masses of molecular gas (Solomon & van den Bout 2005) and high star formation rates (Lutz et al. 2007). However, observations of more typical high-z QSOs (selected without regard to their far-IR properties) do not reveal similarly large gas masses or star formation rates (Maiolino et al. 2007a,b).

Investigations of high-redshift FR II radio galaxies gave us our first view of galaxies at high redshift (e.g. Chambers et al. 1996). As a class, they appear to be very massive systems - the progenitors of present-day giant ellipticals (e.g. Lilly 1989; Willott et al 2003). Their properties imply a high redshift of formation. The fraction of these systems that are detected in rest-frame far-IR rises steeply above redshifts of 2.5 to 3 (Archibald et al. 2001; Reuland et al. 2004). If this emission is powered by star-formation, this may signal the epoch of formation of these systems (Willott et al. 2002; 2003).

The largest and best-studied sample of star forming galaxies at $z > 2$ are the (rest)-UV-selected Lyman Break Galaxies (e.g. Giavalisco 2002), and the related BX and BM samples (Steidel et al. 2004). These have the advantage that the amount of dust-obscuration is modest so that we have a relatively clean view. Based on the general lack of either hard X-ray emission (Laird et al 2006) or of an AGN signature in their rest-UV spectra (Steidel et al. 2002) only a small minority have bright AGN (although see Groves et al. 2007 for a possible caveat). Certainly in these galaxies it appears that the star formation rate is generally much greater than a thousand times the black hole accretion rate.

Another UV-selected population of high-z galaxies are those detected by their Ly$\alpha$ emission-lines (e.g. Rhoads et al. 2000). These galaxies appear to be undergoing significant star formation (with Ly$\alpha$ emission due to photoionization by O stars). They are not detected even in deep stacked X-ray (Wang et al. 2004) or radio continuum (Carilli et al. 2007b) data, and so do not seem to be harboring rapidly growing black holes.

There are several multicolor techniques used to select high-z galaxies in the rest-frame optical. Kriek et al. (2007) have obtained near-IR spectra of a complete sample of 20 K-band-selected galaxies at $z \sim 2.3$. They find that these are massive galaxies (few times $10^{11}$ solar masses). Of the sample, 45% are quiescent (no star formation or AGN), 35% are star-forming galaxies, and 25% are Type 2 AGN. Comparing these demographics to the properties of lower mass systems at the same redshift, they find that AGN activity is preferentially occurring in the most massive galaxies (consistent with a picture of cosmic downsizing). However, in this small sample they see no correlation between the presence of an AGN and star formation.

Moving to longer wavelengths, Spitzer observations of the high-z universe in the rest-frame mid-IR have led to the discovery of a population of highly luminous and dusty systems. The high level of obscuration makes it difficult to robustly separate out the energetic contribution of an AGN and starburst. Mid-IR spectroscopy with Spitzer shows that this is a heterogeneous population, ranging from objects dominated by AGN to objects dominated by starbursts to objects that are clearly starburst-AGN composites (Brand et al. 2007; Sajina et al 2007; Yan et al. 2007). Daddi et al. (2007) find that roughly a quarter of these mid-IR selected galaxies show both a mid-IR excess attributable to an AGN and have X-ray properties consistent with a heavily obscured (Compton-thick) AGN. Their high space density would make them a very significant population of high-z AGN.

At yet longer wavelengths, the sub-mm galaxies at high-z are extremely dusty systems selected on the basis of their rest-frame far-IR emission. Radio continuum observations (Chapman et al. 2001) suggest that the bulk of the far-IR emission is powered by star formation (since they roughly obey the local relation between the radio and far-IR emission defined by local star forming galaxies). This is consistent with inferences based on mid-IR spectroscopy (Valiante et al. 2007; Pope et al. 2007). Most are detected in the hard X-ray band, but the X-ray luminosities imply that the AGN contributes only about 10% of the bolometric luminosity (Alexander et al. 2005). This ratio agrees expectations for an object in which the star formation rate is $\sim 10^3$ times the black hole accretion rate (see above). Thus, these are excellent candidates for witnessing the co-formation of high-mass galaxies and their black holes.

In the next decade, I believe an ambitious program with our new facilities will enable us to make enormous progress in determining the relationship between black hole growth in the early universe. The key will be to undertake panchromatic spectroscopic investigations of large and complete samples of galaxies and AGN selected in a careful and complementary way.

Spectroscopy is the best way to robustly determine the black hole accretion rates and star formation rates. X-ray spectroscopy (IXO) can provide clear evidence of an AGN even when the AGN itself is hidden behind Compton-thick material along our line-of-sight (e.g. Fabian et al. 2000; Levenson et al. 2006). For relatively unobscured objects, spectra in the rest-frame ultraviolet with GSMT and ELT can directly detect the spectroscopic signature of young stars and reflected light from a hidden AGN. Rest-frame optical spectra with JWST can utilize the same diagnostics of star formation and black hole accretion at high-redshift that have been effectively exploited at low redshift. The mid-IR region (JWST with MIRI) is also rich in spectroscopic diagnostics (e.g. Genzel et al. 1998). Heating by the hard radiation field of an AGN has characteristic effects on the molecular gas that can be investigated spectroscopically with ALMA (Carilli et al. 2007a).

The power of large and complete samples to address these issues could be realized by using WFMOS to undertake a spectroscopic survey in the rest-frame UV of a million high-redshift galaxies. This would be a nice side benefit from the core WFMOS science program that aims to constrain Dark Energy through the measurement of baryon acoustic oscillations in the galaxy power spectrum at high redshift.

## *4.4. How are Black Holes Fueled?*

At low redshift AGN activity seems to be triggered by both the inflow of relatively cold gas in medium-mass systems (the Seyfert galaxies) and of hot gas in the most massive galaxies (the radio galaxies). Models imply that the relative importance of the "cold" mode will increase at high redshifts and dominate in both the low and high mass galaxies (e.g. Keres et al 2005; Croton et al 2006). In such models, cold accretion occurs episodically, and many models link the formation of black holes to major merger events (e.g. Hopkins et al. 2007; Di Matteo et al. 2007). Can we see clear evidence at high-z that the growth of black holes is indeed driven by mergers or major accretion events?

Imaging the host galaxies of high-z QSO's is very challenging, even with HST. The investigations to date (Ridgway et al. 2001; Kukula et al 2001; Peng et al. 2006) have determined only the most basic properties (luminosity and size) of the host galaxies for relatively small samples. The images are not sensitive enough to detect large-scale low-surface brightness tidal features while the inner regions are strongly contaminated by the QSO.

In the absence of a bright central QSO, it is easier to look for morphological evidence for mergers in Type 2 AGN. Imaging of high-redshift FR II radio galaxies do show spectacular examples of what appear to be the coalescence of a massive galaxy (e.g. Miley et al. 2006), but these are exceedingly rare systems compared to typical high-z AGN. Based on HST imaging, Pope et al. (2005) and Chapman et al. (2003) conclude that high-z sub-mm galaxies are larger and more asymmetric than other galaxies at these redshifts, and have the complex irregular morphologies suggestive of mergers.

Progress in the next decade will come through two complementary approaches. First, the high angular resolution, stable point spread function, and superb sensitiv-ity of NIRCAM on JWST will enable us to accurately characterized the morphologies of large samples of the host galaxies of high-redshift AGN, including QSOs. This will allow us to quantify the incidence rates of recent mergers in these objects. We may be able to learn about the relative sequencing of the star formation and black hole growth by comparing samples of systems that appear to be in the early vs. late stage of the merger.

Second, the IFUs on JWST, GSMT, and ELT will make it possible to map the ionized gas in the AGN host galaxies, while ALMA will do the same for the molecular gas. These data could provide direct kinematic evidence that we are witnessing the aftermath of a major accretion/merging event and would pinpoint the location and kinematics of the molecular gas that is fueling the star-formation traced by the ionized gas. Programs with the current generation of facilities give us tantalizing examples of the power of such an approach (e.g. Forster Schreiber et al. 2006; Bouche et al. 2007; Nesvadba et al. 2007; Law et al. 2007).

## *4.5. AGN Feedback*

The co-moving rate of accretion onto black holes at z ~2 to 3 was about a factor of ~30 higher than today. Feedback related to this black hole building could have a dramatic effect on the formation and evolution of massive galaxies. Do we see it in action? Here, the situation is somewhat similar to what we see in the modern universe: the most direct and convincing evidence of feedback that is being driven by the AGN and is having a dramatic impact on gas on galactic-scales is found in the form of radio jets.

The presence of galaxy-scale emission-line nebulae around radio-loud QSOs and powerful FR II radio galaxies at high-z has long been known (e.g. McCarthy et al. 1996; Heckman et al 1991a,b). Long-slit spectroscopy of these systems shows the presence of high gas velocities (of-order $10^3$ km/s), strongly suggesting the outflow of substantial amounts of gas. The morphological connection between the ionized gas and the radio sources supports the idea that the outflow is driven by the radio source itself (not by radiation pressure or a spherical wind blown by the black hole). This is consistent with the much smaller, fainter, and more quiescent emission-line nebulae seen around radio-quiet QSOs at the same high redshifts (Christensen et al. 2006).

Recently, very detailed maps in the rest-frame optical have been made for several of the high-z FR II radio galaxies (Nesvadba et al. 2006 and private communication) using near-IR IFU spectroscopy. These data provide for the first time detailed maps of the kinematics and and physical conditions in the regions, and provide convincing confirmation of the idea that the radio sources are likely to be blasting away the entire gaseous halo around the galaxy. This is very exciting, but we must keep in mind that black holes with radio sources this powerful constitute only a small minority (~0.1%) of the rapidly growing supermassive black holes at these redshifts.

With the enormous increase in capability provided by the IFUs on JWST, GSMT, and ELT it would be relatively easy to undertake detailed investigations like this of large and complete samples of all the important classes of AGN at high redshift. The capabilities for high resolution X-ray spectroscopy provided by IXO can be used to investigate the physics of AGN feedback traced by the hot gas. With GSMT and ELT it will possible to undertake high S/N spectroscopy with high spectral resolution for moderately large, complete samples of QSOs to try to directly measure the outflow rates implied by their blueshifted absorption-lines (Korista et al. 2008).

## 5. Final Thoughts

I began this paper by listing all the unanswered questions we have about how the co-evolution of black holes and galaxies actually works. How does gas get into galaxies in the first place? Once inside, how is it transported all the way to the black hole's accretion disk? What astrophysics sets the mass ratio of the gas turned into stars compared to the gas accreted by the black hole (why is this ratio ~$10^3$)? What is the real astrophysics of the feedback from the supermassive black hole that seems to be a crucial ingredient in galaxy evolution? What is the sequencing, both in the sense of overall cosmic history and in the life of an individual galaxy? Does the black hole or galaxy form first? Is the rapid growth of a black hole a once-in-a-lifetime transformative event in the life of a galaxy (e.g. a major merger followed by catastrophic feedback), or is it a more gradual, intermittent process? Does the answer depend on redshift and/or black hole mass?

I have tried to describe our current state of knowledge and ignorance, and have also tried to summarize how I think we can use the amazing new observatories of the next decade to best answer these questions. Given that we are trying to understand a complex cosmic eco-system consisting of hot gas, cold dusty gas, stars, and black holes, it seems clear to me that a panchromatic approach is essential. JWST, ALMA, GSMT/ELT, and IXO will all play crucial roles. I have also tried to emphasize the importance of spectroscopy and of 3-D imaging-spectroscopy in particular. Again, the capabilities of these observatories are a superb match to the problem at hand. Finally, I have emphasized the importance of large surveys of complete and carefully selected samples. The increase in raw sensitivity provided by the new observatories will help make this approach feasible. The huge gain in discovery power provided by of the next generation of multi-object optical and near-IR spectrographs (such as WFMOS) on existing 8 and 10-meter class telescopes will allow us to undertake SDSS-scale surveys at redshifts of one and beyond.

# References

Alexander, D.M., Bauer, F.E., Chapman, S.C., Smail, I., Blain, A.W., Brandt, W.N., & Ivison, R.J 2005, ApJ, 632, 736

Allen, S.W., Dunn, R.J.H., Fabian, A.C., Taylor, G.B., & Reynolds, C.S. 2006, MNRAS, 372, 21

Archibald, E.N.; Dunlop, J.S.; Hughes, D.H.; Rawlings, S.; Eales, S.A.; & Ivison, R.J. 2001, MNRAS, 323, 417

Baldry, I.K., Glazebrook, K., Brinkman, J., Ivezic, Z., Lupton, R., Nichol, R., & Szalay, A. 2004, ApJ, 600, 681

Barden, M. et al. 2005, ApJ, 635, 959

Barger, A.J., Cowie, L.L., Mushotzky, R.F., Yang, Y., Wang, W.-H., Steffen, A.T., & Capak, P. 2005, AJ, 129, 578

Bell, E. F., Wolf, C., Meisenheimer, K., Rix, H.-W., Borch, A., Dye, S., Kleinheinrich, M., Wisotzki, L., McIntosh, D. H. 2004, ApJ, 608, 752

Best, P.N.,Kauffmann, G., Heckman, T. M., & Ivezić, Ž. 2005a, MNRAS, 362, 9

Best, P. N., Kauffmann, G., Heckman, T. M., Brinchmann, J., Charlot, S., Ivezić, Ž., & White, S. D. M. 2005b, MNRAS, 362, 52

Best, P. N., Kaiser, C. R., Heckman, T. M., & Kauffmann, G. 2006, MNRAS, 368, L67

Best, P. N., von der Linden, A., Kauffmann, G.;,Heckman, T. M., & Kaiser, C. R. 2007, MNRAS, 379, 894

Bîrzan, L., Rafferty, D. A., McNamara, B. R., Wise, M. W., & Nulsen, P. E. J. 2004, ApJ, 607, 800

Borys, C., Smail, Ian, Chapman, S. C., Blain, A. W., Alexander, D. M., & Ivison, R. J. 2005, ApJ, 635, 853

Bouche, N. et al. 2007, ApJ, 671, 303

Bower, R, Benson, A., Malbon, R., Helly, J., Frenk, C., Baugh, C., Lacey, C. 2006, MNRAS, 370, 645

Brand, K., Weedman, D., Desai, V., Le Floc'h, E., Armus, L.,, Dey, A., Houck, J. R., Jannuzi, B. T., Smith, H. A.; & Soifer, B. T. 2007, astro-ph:07093119

Braatz, J. A., Henkel, C., Greenhill, L. J., Moran, J. M., & Wilson, A. S. 2004, ApJ, 617, L29

Brinchmann, J., Charlot, S.;,White, S. D. M., Tremonti, C., Kauffmann, G., Heckman, T., & Brinkmann, J. 2004, MNRAS, 351, 1151

Carilli, C. L., Walter, F., Wang, R., Wootten, A., Menten, K., Bertoldi, F., Schinnerer, E., Cox, P., Beelen, A., & Omont, A. 2007a, astro-ph:0703799

Carilli, C., et al. 2007b, ApJS, 172, 518

Chambers, K. C., Miley, G. K., van Breugel, W. J. M., Bremer, M. A. R., Huang, J.-S., & Trentham, N. A. 1996, ApJS, 106, 247

Chapman, S. C.; Richards, E. A.; Lewis, G. F.; Wilson, G.; Barger, A. J. 2001, ApJ, 548, 147

Chapman, S. C., Windhorst, R., Odewahn, S., Yan, H., & Conselice, C. 2003, ApJ, 599, 92

Christensen, L., Jahnke, K., Wisotzki, L., Sánchez, S. F. 2006, A&A, 459, 717

Churazov, E., Brüggen, M., Kaiser, C. R., Böhringer, H., & Forman, W. 2001, ApJ, 554, 261
Cid Fernandes, R., Heckman, T., Schmitt, H., Delgado González. R, Storchi-Bergmann, T., 2001, ApJ, 558, 81
Croton, D. J., Springel, V., White, S. D. M., De Lucia, G., Frenk, C. S., Gao, L., Jenkins, A., Kauffmann, G., Navarro, J. F., Yoshida, N. 2006, MNRAS, 365, 11
Daddi, E. et al. 2007, ApJ, 670, 173
Davies, R. I., Genzel, R., Tacconi, L.,Mueller Sánchez, F., & Sternberg, A. 2006, astroph: 0612009
Davies, R. I., Mueller Sánchez, F., Genzel, R., Tacconi, L. J., Hicks, E. K. S., Friedrich, S., & Sternberg, A. 2007 ApJ, 671, 1388
Dunlop, J. S., & Peacock, J. A. 1990, MNRAS, 247, 19
Di Matteo, T., Colberg, J., Springel, V., Hernquist, L.,; & Sijacki, D. 2007, astroph: 07052269
Erwin, P, & Sparke, L.S. 2002, AJ, 124, 65
Faber, S. et al. 2007, ApJ, 665, 265
Fabian, A., Iwasawa, K, Reynolds, C.S., & Young, A.J. 2000, PASP, 112, 1145
Fabian, A., Sanders, J., Taylor, G., Allen, S., Crawford, C., Johnstone, R., Iwasawa, K. 2006, MNRAS, 366, 417
Fanaroff, B. L., & Riley, J. M. 1974, MNRAS, 167, 31
Fathi, K., Storchi-Bergmann, T., Riffel, Rogemar, A., Winge, C., Axon, D. J., Robinson, A., Capetti, A., & Marconi, A. 2006, ApJ, 641, L25
Ferrarese, L. & Merritt, D. 2000, ApJ, 539, L9
Fine, S. et al. 2006, MNRAS, 373, 613
Forster Schreiber, N. et al. 2006, AJ, 131 1891
Gebhardt, K. et al. 2000, ApJ, 539, L13
Genzel, R. et al. 1998, ApJ, 498, 579
Genzel, R., & Karas, V. 2007, astro-ph:07041281
Giavalisco, M. 2002, ARA&A, 40, 579
González Delgado, R.., Cid Fernandes, R. Pérez, E., Martins, L. P. Storchi-Bergmann, T., Schmitt, H., Heckman, T., Leitherer, C. 2004, ApJ, 605, 127
Granato, G. L., Silva, L., Monaco, P., Panuzzo, P., Salucci, P., De Zotti, G., Danese, L. 2001, MNRAS, 324, 757
Greene, J., and Ho, L. 2005, ApJ, 627, 721
Grogin, N.A., et al 2005, ApJ, 627, L97
Groves, B., Heckman, T., & Kauffmann, G. 2006, MNRAS, 371, 1559
Hao. L. et al. 2005, AJ, 129, 1783
Häring, N,, & Rix, H.-W. 2004, ApJ, 604, L89
Hasinger, G., Miyaji, T., & Schmidt, M. 2005, A&A, 441, 417
Heckman, T. M., Miley, G. K., Lehnert, M. D., & van Breugel, W. 1991a, ApJ, 370, 78
Heckman, T. M., Lehnert, M. D., Miley, G. K., & van Breugel, W. 1991b, ApJ, 381, 373
Heckman, T., Kauffmann, G., Brinchmann, J., Charlot, S., Tremonti, C., & White, S. 2004, ApJ, 613, 109
Hicks, E., & Malkan, M. 2008, ApJS, 174, 31
Hopkins, P. F., Robertson, B., Krause, E., Hernquist, L., & Cox, T. J. 2006, ApJ, 652, 107
Hopkins, P. F., Hernquist, L. Cox, T. J., & Keres, D. 2007, astro-ph:07061243
Jogee, S. et al. 2004, ApJ, 615, L105
Kauffmann, G., & Haehnelt, M 2000, MNRAS, 311, 576
Kauffmann, G., White, S.D.M, & Guiderdoni, B. 1993, MNRAS, 264, 201
Kauffmann, G., et al 2003a, MNRAS, 341, 33

Kauffmann, G., Heckman, T. M., White, S. D. M., Charlot, S., Tremonti, C., , Peng, E. W., Seibert, M., Brinkmann, J., Nichol, R. C., SubbaRao, M., & York, D. 2003b, MNRAS, 341, 54

Kauffmann, G. et al. 2003c, MNRAS, 346, 1055

Kauffmann, G., White, S. D. M., Heckman, T. M. Ménard, B., Brinchmann, J.,; Charlot, S., Tremonti, C., & Brinkmann, J. 2004, MNRAS, 353, 713

Kauffmann, G. et al. 2007a, ApJS, 173, 357

Kauffmann, G., Heckman, T., Best, P. 2007b, astro-ph: 07092911

Kaspi, S., Maoz, D., Netzer, H., Peterson, B.M., Vestergaard, M., & Jannuzi, B. T. 2005, 629, 61

Kereš, D., Katz, N., Weinberg, D. H., & Davé, R. 2005, MNRAS, 363, 2

Korista, K., Bautista, M., Arav, N., Moe, M., Constantini, E., & Benn, C. 2008, astro-ph: 0807230

Kormendy, J. & Kennicutt, R. 2004, ARA&A, 42, 603

Knapen, J, Schlosman, I, & Peletier, R. 2000, ApJ, 529, 97

Kriek, M. et al. 2007, ApJ, 669, 776

Krolik, J.H., & Begelman, M. 1988, ApJ, 329, 702

Krongold, Y., Nicastro, F., Elvis, M., Brickhouse, N., Binette, L., Mathur, S., & Jiménez-Bailón, E. 2007, ApJ, 659, 1022

Kukula, M. J., Dunlop, J. S., McLure, R. J., Miller, L., Percival, W. J., Baum, S. A., & O'Dea, C. P. 2001, MNRAS, 326, 1533

Laine, S., Shlosman, I., Knapen, J. H., & Peletier, R. F. 2002, ApJ, 567, 97

Laird, E. S., Nandra, K., Hobbs, A., & Steidel, C. C. 2006, MNRAS, 373, 217

Lauer, T. R., Tremaine, S., Richstone, D., & Faber, S. M. 2007, ApJ, 670, 249

Law, D. R., Steidel, C. C., Erb, D. K., Larkin, J. , Pettini, M., Shapley, A.,, & Wright, S. A. 2007, ApJ, 669, 929

Levenson, N. A., Heckman, T. M., Krolik, J. H., Weaver, K. A., & Życki, P. T. 2006, ApJ, 648, 111

Li, C., Kauffmann, G., Heckman, T. M., White, S. D. M., & Jing, Y. P. 2007, astro-ph:07120383

Lilly, S. 1989, ApJ, 340, 77

Lindt-Krieg, E., Eckart, A., Neri, R. Krips, M., Pott, J. -U., Garcia-Burillo, S., Combes, F. 2007, astro-ph:07123133

Lutz, D., Sturm, E., Tacconi, L. J., Valiante, E., Schweitzer, M., Netzer, H., Maiolino, R., Andreani, P., Shemmer, O., & Veilleux, S. 2007, ApJ, 661, L25

Maiolino, R., Shemmer, O., Imanishi, M., Netzer, H., Oliva, E., Lutz, D., & Sturm, E.,. 2007a, A&A, 468, 979

Maiolino, R. et al., 2007b, A&A, 472, L33

Marconi, A., & Hunt, L.K. 2003, ApJ, 589, L21

Marconi, A., Risaliti, G., Gilli, R., Hunt, L. K., Maiolino, R., & Salvati, M. 2004, MNRAS, 351, 169

Martin, D.C., et al. 2007, ApJS, 173, 342

Martini, P., Regan, M. W., Mulchaey, J. S., & Pogge, R. W. 2003, ApJ, 589, 774

McCarthy, P. , Baum, S., & Spinrad, H. 1996, ApJS, 106, 281

McLure, R. J., Jarvis, M. J., Targett, T. A., Dunlop, J. S., Best, P. N. 2006, MNRAS, 368, 1395

McNamara, B.R., & Nulsen, P.E.J. 2007, ARA&A, 45, 117

Mihos, J.C., & Hernquist, L. 1994, 425, L13

Miley, G.K., et al. 2006, ApJ, 650, L29

Moran, J., Humphreys, L., Greenhill, L., Reid, M., & Argon, A. 2007, astro-ph:07071032

Mulchaey, J. & Regan, M. 1997, ApJ, 482, L135

Nandra, K., Georgakakis, A., Willmer, C. N. A., Cooper, M. C., Croton, D. J., Davis, M., Faber, S. M., Koo, D. C., Laird, E. S., & Newman, J. A. 2007, ApJ, 660, L11

Nelson, C. & Whittle, M. 1996, ApJ, 465, 96

Nesvadba, N. P. H., Lehnert, M. D.;,Eisenhauer, F., Gilbert, A., Tecza, M., Abuter, R. 2006, ApJ, 650, 693

Nesvadba, N. P. H., Lehnert, M. D., Davies, R. I., Verma, A., & Eisenhauer, F. 2007, astro-ph:07111491

Peng, C. Y., Impey, C. D., Rix, H.-W., Kochanek, C. S., Keeton, C. R., Falco, E. E., Lehár, J., & McLeod, B. A. 2006, ApJ, 649, 616

Pierce, C.M. et al. 2007, ApJ, 660, L19

Pope, A., Borys, C., Scott, D., Conselice, C., Dickinson, M., & Mobasher, B. 2005, MNRAS, 358, 149

Pope, A., Chary, R.-R., Alexander, D. M., Armus, L., Dickinson, M., Elbaz, D.,; Frayer, D., Scott, D., & Teplitz, Harry 2007, astro-ph:70111553

Reichard, T., Heckman, T. Rudnick, G., Brinchmann, J., Kauffmann, G., & Wild, V. 2008, in preparation

Reuland, M., Röttgering, H., van Breugel, W. & De Breuck, C. 2004, MNRAS, 353, 377

Rhoads, J. E., Malhotra, S., Dey, A., Stern, D., Spinrad, H., & Jannuzi, B. T. 2000, ApJ, 545, L85

Ridgway, S. E., Heckman, T. M., Calzetti, D., & Lehnert, M. 2001, ApJ, 550, 122

Riffel, R., Storchi-Bergmann, T., Winge, C., Beck, T., & Schmitt, H. 2008, MNRAS, 385, 1129

Rovilos, E., & Georgantopoulos, I. 2007, A&A, 475, 115

Rovilos, E., Georgakakis, A., Georgantopoulos, I., Afonso, J., Koekemoer, A. M., Mobasher, B., & Goudis, C. 2007, A&A, 466, 119

Sadler , E. et al. 2007, MNRAS, 381, 211

Sajina, A., Yan, L., Armus, L., Choi, P., Fadda, D., Helou, G., & Spoon, H. 2007, ApJ, 664, 713

Salviander, S., Shields, G. A., Gebhardt, K., & Bonning, E. W. 2007, ApJ, 662, 131

Schiminovich, D et al 2007, ApJS, 173, 315

Shields, G. A., Gebhardt, K., Salviander, S., Wills, B. J., Xie, B., Brotherton, M. S., Yuan, J., Dietrich, M. 2003, ApJ, 583, 124

Simões Lopes, R. D., Storchi-Bergmann, T., de Fátima Saraiva, M., & Martini, P. 2007, ApJ, 665, 718

Solomon, P. M., & Vanden Bout, P. A. 2005, ARA&A, 43, 677

Spergel, D. et al. 2007, ApJS, 170, 377

Steidel, C. C., Hunt, M. P., Shapley, A E., Adelberger, K. L., Pettini, M., Dickinson, M., & Giavalisco, M. 2002, ApJ, 576, 653

Steidel, C. C., Shapley, Al. E., Pettini, M., Adelberger, K. L,. ,Erb, Dawn K., Reddy, N A., Hunt, M. P. 2004, ApJ, 604, 534

Storchi-Bergmann, T., Dors, O.L., Jr., Riffel, R. A., Fathi, K., Axon, D. J., Robinson, A.,; Marconi, A., & Östlin, G. 2007, ApJ, 670, 959

Tegmark, M., et al. 2004, Phys. Rev. D 69, 3051

Tremaine, S. et al. 2002, ApJ, 574, 740

Tremonti, C. A., Moustakas, J., & Diamond-Stanic, A. M. 2007, ApJ, 663, L77

Tristram, K.R.W. et al. 2007, A&A, 474, 837

Ueda, Y., Akiyama, M., Ohta, K., & Miyaji, T. 2003, ApJ, 598, 886

Valiante, E., Lutz, D., Sturm, E., Genzel, R., Tacconi, L. J., Lehnert, M. D., Baker, A. J. 2007, ApJ, 660,1060

Veilleux, S., Cecil, G., & Bland-Hawthorn, J. 2005, ARA&A, 43, 769

Vestergaard, M., Fan, X., Tremonti, C. A., Osmer, P. S., & Richards, G.. T. 2008, astro-ph:0801243
Wang, J. X., Rhoads, J. E., Malhotra, S., Dawson, S., Stern, D., Dey, A., Heckman, T. M., Norman, C. A. & Spinrad, H. 2004, ApJ, 608, L21
Weymann, R., Carswell, R., & Smith, M. 1981, ARA&A, 19, 41
Wild, V., Kauffmann, G., Heckman, T., Charlot, S., Lemson, G., Brinchmann, J., Reichard, T., & Pasquali, A. 2007, MNRAS, 381, 543
Willott, C. J., Rawlings, S., Archibald, E. N., & Dunlop, J. S. 2002, MNRAS, 331, 435
Willott, C. J., Rawlings, S., Jarvis, M. J., & Blundell, K. M. 2003, MNRAS, 339, 173
Yan, L., Sajina, A., Fadda, D., Choi, P., Armus, L., Helou, G., Teplitz, H., Frayer, D., & Surace, J. 2007, ApJ, 658, 778